\pdfoutput=1 
\documentclass[%
 aip,
 rsi,
 amsmath,amssymb,
reprint,%
]{revtex4-2}

\usepackage{graphicx}
\usepackage{dcolumn}
\usepackage{bm}

\usepackage[utf8]{inputenc}
\usepackage[T2A,T1]{fontenc}
\usepackage{mathptmx}
\usepackage[russian,english]{babel}

\usepackage{upgreek}

\usepackage{adjustbox}

\draft 

\def\Rs{R_\mathrm{s}}

\def\Cp{C_\mathrm{p}}
\def\Vd{V_\mathrm{d}}
\def\Id{I_\mathrm{d}}
\def\Vin{V_\mathrm{in}}
\def\I2C{I$^2$C}

\begin{document}

\title{Stabilizing amplifier with a programmable load line for characterization of nanodevices with negative differential resistance}

\author{T. Hennen}
 \affiliation{IWE II, RWTH Aachen University, 52074 Aachen, Germany}
\author{E. Wichmann}
 \affiliation{IWE II, RWTH Aachen University, 52074 Aachen, Germany}
\author{R. Waser}
 \affiliation{IWE II, RWTH Aachen University, 52074 Aachen, Germany}
\author{D. J. Wouters}
 \affiliation{IWE II, RWTH Aachen University, 52074 Aachen, Germany}
\author{D. Bedau}
  \email{daniel.bedau@wdc.com}
  \affiliation{Western Digital San Jose Research Center, 5601 Great Oaks Pkwy, San Jose, CA 95119}

\date{\today}

\begin{abstract}
Resistive switching devices and other components with negative differential
resistance (NDR) are emerging as possible electronic constituents of
next-generation computing architectures. Due to the NDR effects exhibited,
switching operations are strongly affected by the presence of resistance in
series with the memory cell. Experimental measurements useful in the development
of these devices use a deliberate addition of
series resistance, which can be done either by integrating resistors on-chip or
by connecting external components to the wafer probing system. The former
approach is considered inflexible because the resistance value attached to a
given device cannot be changed or removed, while the latter approach tends to
create parasitic effects that impact controllability and interfere with
measurements. In this work we introduce a circuit design for flexible
characterization of two-terminal nanodevices that provides a programmatically
adjustable external series resistance while maintaining low parasitic
capacitance. Experimental demonstrations are given that
show the impact of the series resistance on NDR and resistive switching
measurements.
\end{abstract}


\maketitle

\section{Introduction}\label{ch:introduction}
Increasing research and development efforts aim to produce new types of scalable
two-terminal nanodevices suitable for storing and processing information in both
traditional and neuromorphic computing architectures
\cite{burr_neuromorphic_2017, li_resistive_2017, sangwan_neuromorphic_2020}.
Emerging devices are often based on incompletely understood mechanisms, and may
exhibit strong non-linearity, negative differential resistance (NDR),
oscillations, stochasticity, and memory effects. In assessing the electrical
capabilities of resistive switching devices such as
ReRAM\cite{waser_redox-based_2009}, it is important to consider not only the
device material properties but also the effects of feedback, runaway, excess
electrical stresses, and the general role of the driving circuitry on
measurement data.

Electrical measurements of patterned devices are inevitably carried out in the
presence of resistance in series with the active material volume of the cell.
This series resistance, commonly of unknown value
\cite{hardtdegen_internal_2016, ibanez_non-uniform_2021}, may originate from a
combination of the electrode leads, inactive layers of the material stack, or
the triode region of a series FET current limiter. Internal and external series
resistance adds current-voltage feedback to the system that affects stability
and influences the operational behavior in important ways.
Modification of switching speeds, threshold voltage/currents, and the range of
achievable resistance states have all been observed and discussed theoretically
\cite{fantini_intrinsic_2012, hennen_switching_2019, gonzalez_current_2020,
maldonado_experimental_2021, hardtdegen_improved_2018, strachan_state_2013}.

A series resistance is often intentionally placed to play
the necessary role of an energy limiting mechanism, where its value can mean the difference between a functioning and
non-functioning device. As an experimental technique, it is useful
to be able to place different resistance values in series with the device under
test (DUT) and examine the effect on switching processes. Here, the linearity of
the resistive load is a convenient property for mathematical modelling; The
circuit response of the simple two element series configuration (2R) is easily
predictable through load line analysis in the ideal case, and is also
straightforward to treat analytically in the presence of commonly encountered
parasitics.

\begin{figure}
    \maxsizebox{\columnwidth}{!}{\includegraphics[scale=1]{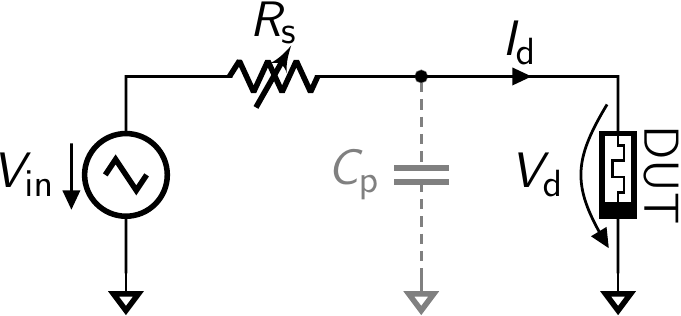}}
    \caption{A simple circuit configuration for device characterization uses a waveform generator and an external resistance in series with the DUT.  In practice, the effect of the parasitic capacitance in parallel with the device requires careful attention.}
    \label{fig:RC_circuit}
\end{figure}

Another advantage of the 2R configuration is ease of implementation relative to
integration of active FET devices on a test chip, with the latter requiring
substantial fabrication cycle time. However, integrating calibrated series
resistances on-chip is inflexible because each cell is attached to a single static
resistance value that cannot be changed or removed. Scenarios often arise that
give good reason to alter or remove the series resistance \textit{in situ}.
Notably, devices possessing steady states with S-type or N-type NDR each have
different criteria for stable characterization, and both types are
present in the SET and RESET processes of ReRAM,
respectively\cite{fantini_intrinsic_2012}. This imposes different requirements
for the series resistance value even within a single switching
cycle.

Where an adjustable series resistance is required, it must be implemented
externally to the wafer. The main practical challenge associated with this is
that parasitic capacitance $\Cp$ at the node shared with the DUT is highly
detrimental and difficult to avoid (Fig. \ref{fig:RC_circuit}). This stray
capacitance slows down the dynamic response of the circuit, degrading the
ability to control and to measure the voltage and current
experienced by the active cell volume versus time. Coupled with rapid
conductance transitions of the DUT, harmful overshoot transients are generated
that strongly impact the observed switching behavior and can cause irreversible
damage \cite{kinoshita_reduction_2008, lu_elimination_2012,
sharma_electronic_2014, meng_temperature_2020}.

While singular through-hole resistors are a common external solution, their use entails manually switching between resistance values where required. However, the stochastic nature of resistive switching cells is such that they benefit greatly from a statistical treatment using automated measurements with programmable parameters. In this work we present an external circuit design providing an adjustable linear series resistance for flexible wafer-level device characterization. The circuit, based on a digital potentiometer (digipot) chip, is remotely programmable over USB between 528 resistance levels. Importantly, the voltage signal at the low-capacitance DUT node is directly amplified for synchronous measurement with the DUT current with a bandwidth over 200~MHz.  We demonstrate the circuit operation for automated characterization of NDR devices and for cycling of bipolar ReRAM cells with high speed voltage sweeps.

\section{Design} \label{ch:designPrinciples}

\begin{figure}
    \includegraphics[width=1\columnwidth]{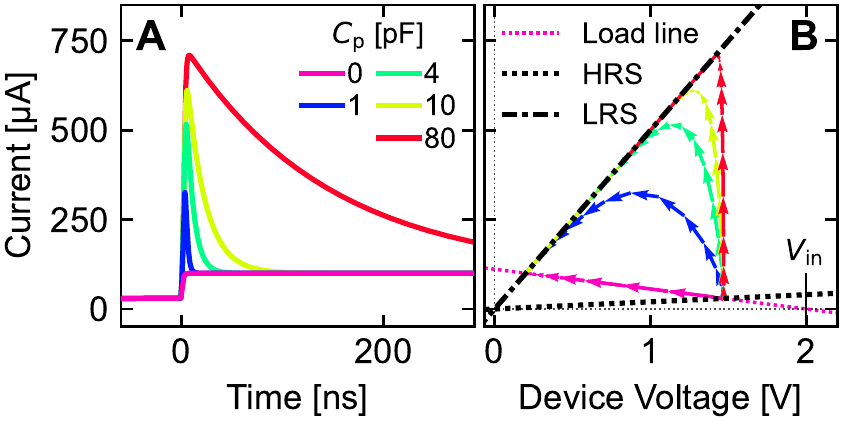}
    \caption{Simulations (using Eq.~\ref{eq:diff_eq}) of $\Id,\Vd$ transients following a rapid resistance transition of the DUT with $\Vin = 2$~V and different values of $\Cp$. Subplot \textbf{(A)} shows $\Id$ vs. $t$ while \textbf{(B)} shows $\Id$ vs. $\Vd$ of the same simulations. The DUT resistance value is assumed to change exponentially in time from a high resistance state (HRS) of $50~\mathrm{k}\upOmega$ to a low resistance state (LRS) of $2~\mathrm{k}\upOmega$ with time constant 1~ns. During and following the transition, the device is subjected to excess currents relative to the load line, an effect which is reduced by using lower $\Cp$ values.}
    \label{fig:overshoot_sim}
\end{figure}

Applying Kirchhoff's current law, the dynamical equation governing the time evolution of the device voltage in the circuit of Fig.~\ref{fig:RC_circuit} is
\begin{align}\label{eq:diff_eq}
    \Cp \frac{d\Vd(t)}{dt} &= \frac{\Vin(t) - \Vd(t)}{\Rs} - \Id(t,~\ldots),
\end{align}
where $t$ is time and $\Id$ in general depends on $\Vd$ and other internal state variables of the DUT. Possible steady state solutions lie on the $\Vd$-nullcline,
\begin{align}\label{eq:load_line}
    \Vd = \Vin - \Id \Rs,
\end{align}
also known as the load line.  For fast conductance switching events that are common in the targeted material systems, transient deviations from the load line occur as seen in the simplified situation of Fig.~\ref{fig:overshoot_sim}. During such transients, the excess energy delivered to the DUT due to capacitive discharge is significant and can strongly influence the end result of the switching process.

While the potential for overshooting transients is unavoidable in the context of
a passive feedback arrangement,
it is important that they are controlled to the extent possible and accurately
measured when they occur. The only way that overshoots can be reduced in the
discussed configuration is by minimizing the value of $\Cp$. Practically this
means that a coaxial cable, acting approximately as a parasitic capacitance of
100~pF/m, cannot be used to connect $\Rs$ to the DUT. The series resistance
should rather be placed as close as possible to the DUT, with the components
carefully selected and the printed circuit board (PCB) layout designed for low
contribution to the total $\Cp$.

High fidelity current measurements can be achieved by amplification of the
voltage across a ground referenced shunt termination following transmission over
a coaxial line. Using this type of current measurement,
positioning the DUT (rather than $R_s$) adjacent to the shunt is generally
preferred because it avoids low pass filtering of the $\Id$ signal, allowing
measurement of $\Id$ at a high bandwidth that is independent of the resistance
state of the device. With prior knowledge of $\Rs$, Eq.~\ref{eq:load_line} is
often used to calculate the $\Vd$ from a measurement of $\Id$ and
$\Vin$, but there are several drawbacks
associated with this method. One is the inaccuracy that comes from neglecting
the capacitive currents of the left-hand side of Eq.~\ref{eq:diff_eq}. Another
problem is measurement noise introduced by the $\Id\Rs$ term, as the small $\Id$
signal with high relative error is multiplied by a potentially large $\Rs$
value. It is therefore advantageous to directly amplify the voltage at the DUT
electrode rather than attempt to calculate it from other measured signals.

\begin{figure}[h]
\centering
\maxsizebox{\columnwidth}{!}{\includegraphics[scale=1]{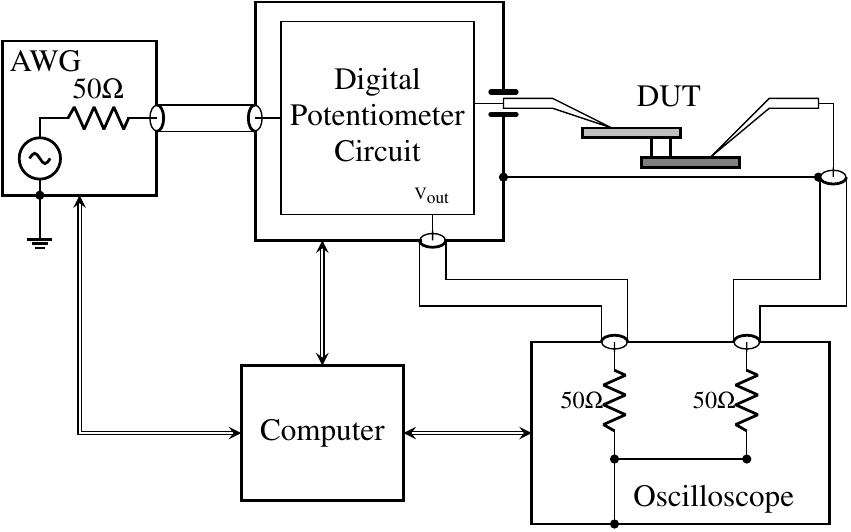}}
	\caption{Schematic depiction of the overall measurement setup.  An arbitrary
    waveform generator (AWG) produces the driving signal $\Vin(t)$, and the
    resulting current is sampled after the right electrode via the 50~$\upOmega$
    shunt of the oscilloscope input. A second oscilloscope channel simultaneously captures the amplified voltage at the left electrode. A ground jumper provides a low inductance return path and reduces RF interference. All instruments are under computer control.
	}
	\label{fig:measurement_setup}
\end{figure}

Following from these considerations, the basic intended configuration of
external instruments and the designed circuit can be seen in
Fig~\ref{fig:measurement_setup}. If sufficient resolution is not obtained by
sampling the current with a bare oscilloscope input, additional voltage
amplification should be placed at the termination, where the use of several output
stages is beneficial for dynamic range. Note that the length of the coaxial
lines for DUT voltage and current sampling should be matched so that
post-processing is not needed for signal synchronization.

A commercial integrated circuit, the DS1808 digipot from Maxim Integrated, was
chosen as the central component to control the series resistance, $\Rs$.
Internally it contains two separate potentiometers, each consisting of a chain
of 32 resistors whose junctions can be connected to a "wiper" output via a set
of CMOS transmission gates (analog switches). For each potentiometer, there are
32 available resistance settings spaced logarithmically (piecewise) from
approximately $300~\upOmega $ to $45~\mathrm{k\upOmega}$. According to the
published specifications, the DS1808 has a maximum parasitic capacitance of
10~pF and a maximum voltage range of $\pm12$~V\cite{DS1808}.

To increase the coverage of $\Rs$ values, the PCB is routed in a way that allows
connection of both potentiometers either in series or in parallel by connecting
or opening solder jumper pads. While a connection to a single potentiometer
remains a possibility, the number of unique settings is increased to 528 between $600~\upOmega - 90~\mathrm{k}\upOmega$ for the series combination and between $150~\upOmega - 22.5~\mathrm{k}\upOmega$ for the parallel combination. Because a low resistance setting below 300~$\upOmega$ is not provided by the digipots, a reed switch was also included on the PCB to add an option to short the input directly to the output.

For amplification of the output voltage, a THS3091 current-feedback operational amplifier from Texas Instruments was used in a non-inverting configuration. This device features low distortion, low noise, a bandwidth of 210 MHz, and a slew rate of $\mathrm{7300 V /\upmu s}$ while adding only 0.1~pF parasitic capacitance\cite{THS3091}.

\begin{figure}[h]
\centering
\maxsizebox{\columnwidth}{!}{\includegraphics[scale=1]{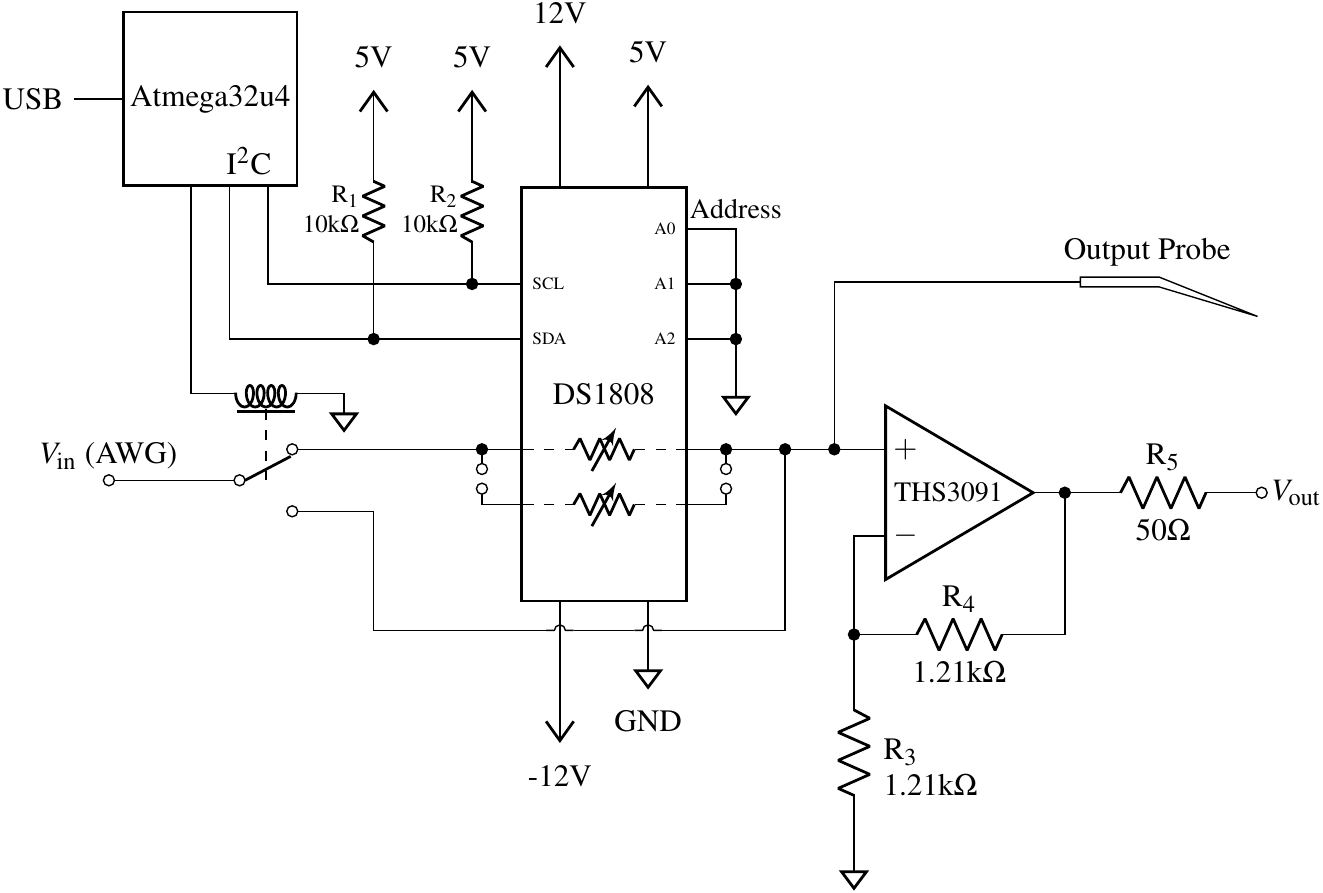}}
\caption{Simplified schematic of the digipot measurement circuit.  An Atmega32u4 microcontroller USB-serial interface communicates to the DS1808 digipot via an \I2C bus. A SPDT reed relay can be actuated in order to bypass the digipot and make a direct connection between input and output. The voltage at the output is amplified by a THS3091 non-inverting follower.}
\label{fig:completeSchematic}
\end{figure}

\begin{figure}[h]
    \setlength{\fboxsep}{0pt}%
    \setlength{\fboxrule}{1pt}%
    \fbox{\includegraphics[width=.8\columnwidth]{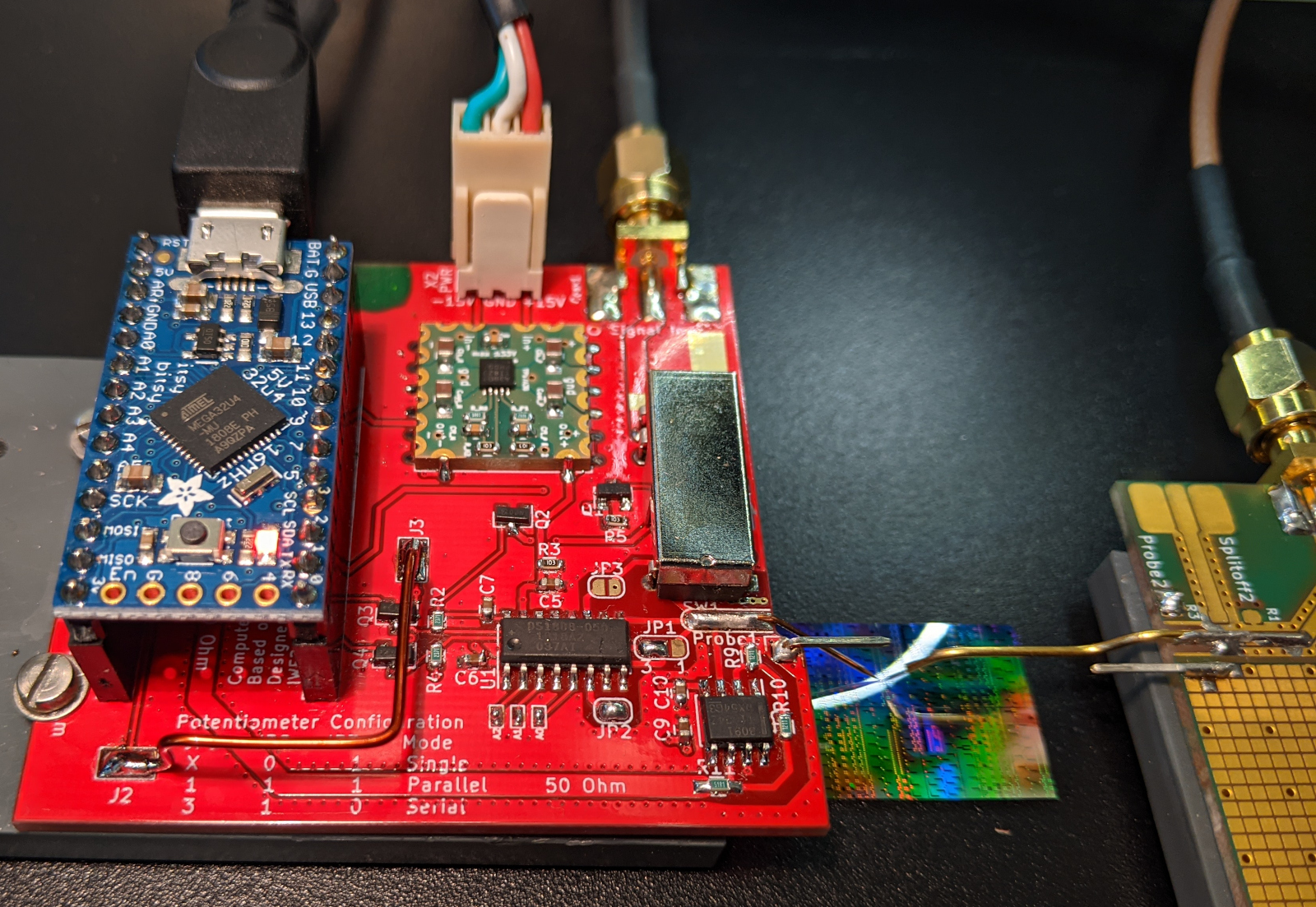}}
	\caption{A photograph of the probing PCB contacting a test chip. A non-coaxial BeCu probe tip is soldered directly to the output of the main PCB (red), which uses SMA connectors for additional input and output signals. An elevated PCB (blue) contains the microcontroller USB interface (Adafruit ItsyBitsy 32u4). A square PCB module (green) functions as a low noise dual voltage regulator providing $\pm$ 12~V to the system. The right probe is directly connected to a 50~$\upOmega$ oscilloscope input.}
    \label{fig:probingsetup}
\end{figure}

All on-board settings are controlled via an Atmega32u4 microcontroller programmed as a USB serial interface to the PC.  Control of the $\Rs$ value is accessible using any programming language able to open a serial COM connection and send a simple command composed of three integer values corresponding to the wiper positions and the state of the bypass relay.  The total time from issuing a serial command to $\Rs$ reaching a new value is limited by USB / \I2C communication, and is typically less than 300~$\upmu$s. The overall circuit design is visualized in the a block diagram of Fig~\ref{fig:completeSchematic}, and a corresponding fabricated PCB is pictured in Fig.~\ref{fig:probingsetup}.

\section{Measurements}

\begin{figure}[h]
    \centering
    \includegraphics[]{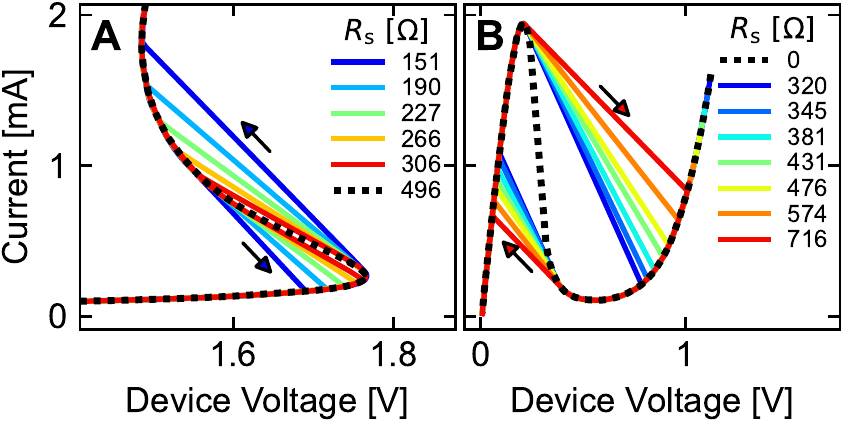}
    \caption{Voltage sweeping measurements of NDR devices using different resistance settings. \textbf{(A)} 90$\times$500$\times$500~nm S-type VCrOx device \cite{hennen_forming-free_2018}, stabilized for $\Rs > 400~\upOmega$. \textbf{(B)} N-type Ga-As tunnel diode 3\foreignlanguage{russian}{И}306E, stabilized for $\Rs = 0~\upOmega$.}
    \label{fig:NDR2}
\end{figure}

For quasistatic measurements of classical NDR materials using a series resistance, saddle-node bifurcations can occur that separate the NDR characteristic into stable and unstable regions. The range of the unstable region is determined by the value of the series resistor, with the bifurcations occurring where the derivative of the NDR curve voltage with respect to current crosses $-\Rs$. While sweeping voltage, sudden current jumps are observed for sufficiently low values of $\Rs$ in S-type NDR (Fig.~\ref{fig:NDR2}A) and for sufficiently high values of $\Rs$ in N-type NDR (Fig.~\ref{fig:NDR2}B). Thus, an adaptable $\Rs$ value allows control of the conditions under which each of these characteristic curves, which contain important information, can be measured.

\begin{figure}[h]
    \centering
    \includegraphics[]{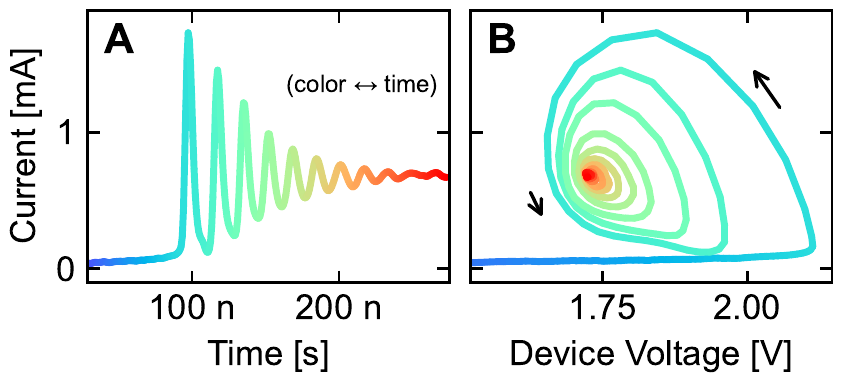}
    \caption{Oscillations (57 MHz) occurring in a 30$\times$250$\times$250~nm \mbox{S-type} VCrOx NDR device\cite{hennen_forming-free_2018} following a square voltage pulse \mbox{$\Vin = 0~V \rightarrow 2.5~V$} using $\Rs = 1083~\upOmega$. With the line color mapped to time of measurement, \textbf{(A)} shows $I_d$ vs $t$ of the transient, and \textbf{(B)} shows the trajectory of the same data on the $\Id,\Vd$ plane.}
    \label{fig:osc}
\end{figure}

Where the material mechanism of NDR is dynamic and reversible, the presence of
$\Cp$ makes the measurement circuit prone to transient oscillations, and stable
oscillatory limit cycles can also occur. Useful in these cases, the presented
circuit is able to capture high speed transients and accurately project them
onto the device $\Id,\Vd$ plane (Fig.~\ref{fig:osc}). This data can be used for
device modelling and circuit simulations, each relevant for example in the
ongoing investigations of coupled oscillatory devices in neuromorphic systems.

\begin{figure}[h]
    \centering
    \includegraphics[width=1\columnwidth]{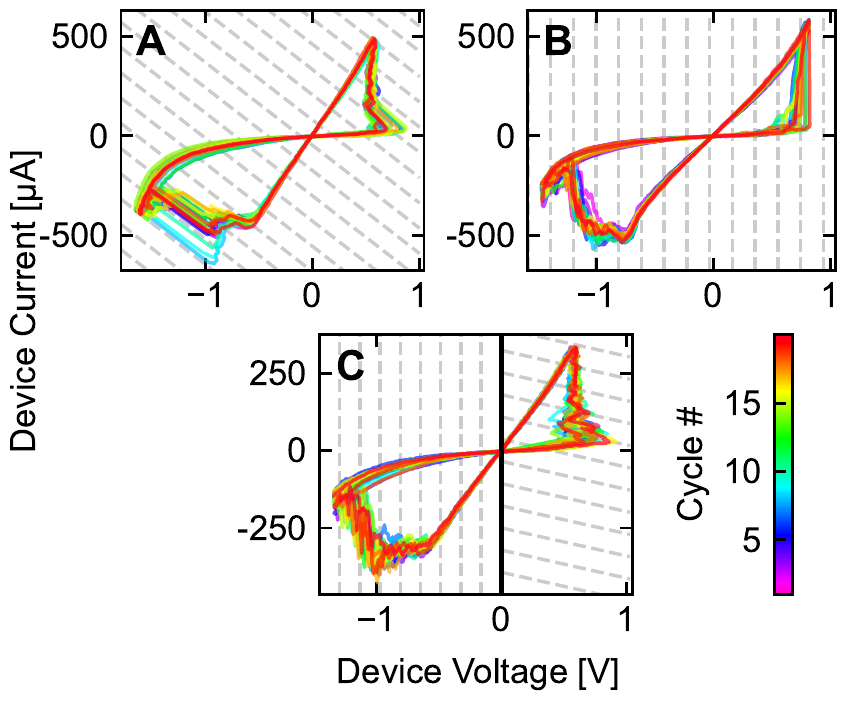}
    \caption{Cycling measurements of a 100 nm ReRAM
    device\cite{hennen_current-limiting_2021} using bipolar triangular voltage
    sweeps with 1~ms duration. Each subplot contains 20 consecutive switching cycles differentiated by color. The value of added series resistance is indicated by the dashed lines with gradient -1/$\Rs$. Transition behavior differs considerably when using \textbf{(A)} 2.4~k$\upOmega$, \textbf{(B)} 0~$\upOmega$, and \textbf{(C)} 11~k$\upOmega$ for positive polarity and 0~$\upOmega$ for negative polarity.}
    \label{fig:reram_measurement}
\end{figure}

In ReRAM devices, NDR behavior is mediated by a combination of Joule heating and
migration of point defects in the oxide material that locally increase its
conductivity\cite{waser_redox-based_2009}. Altering the $\Rs$ value allows these
transitions to be probed in different ways, as seen in the example measurements
of Fig.~\ref{fig:reram_measurement}. In analogy to the NDR measurements of
Fig.~\ref{fig:NDR2}, a fixed value of $\Rs$ can result in sudden and unstable
transitions for one or both of the SET or RESET processes. By switching the
value of $\Rs$ during the measurement (Fig.~\ref{fig:reram_measurement}C) it is
observed that runaway load line transitions can be suppressed by appropriate
selection of the external feedback.

\section{Conclusion}

When performing electrical measurements of resistive switching systems, the use
of well understood circuitry is critical for realistic evaluation. With isolated
devices vulnerable to runaway transitions, a series resistance circuit provides
a simple means for control and tractable analysis of switching processes. In
this context, parasitic capacitance is an important factor, and the values of
both $\Rs$ and $\Cp$ lead to different switching outcomes in general. We have
presented a circuit design for synchronized measurement of switching
trajectories at high speed while using a programmable linear series resistance
with low parasitic capacitance. Using this circuit, possible implications of the
physical processes that accompany runaway transitions can be conveniently
investigated, yielding insights into how optimal control can eventually be
achieved.

\section*{Data Availability}
The data that support the findings of this study are available from the corresponding author upon reasonable request.

\bigskip

\bibliography{bibliography}

\providecommand{\noopsort}[1]{}\providecommand{\singleletter}[1]{#1}%
\begin{thebibliography}{20}%
\makeatletter
\providecommand \@ifxundefined [1]{%
 \@ifx{#1\undefined}
}%
\providecommand \@ifnum [1]{%
 \ifnum #1\expandafter \@firstoftwo
 \else \expandafter \@secondoftwo
 \fi
}%
\providecommand \@ifx [1]{%
 \ifx #1\expandafter \@firstoftwo
 \else \expandafter \@secondoftwo
 \fi
}%
\providecommand \natexlab [1]{#1}%
\providecommand \enquote  [1]{``#1''}%
\providecommand \bibnamefont  [1]{#1}%
\providecommand \bibfnamefont [1]{#1}%
\providecommand \citenamefont [1]{#1}%
\providecommand \href@noop [0]{\@secondoftwo}%
\providecommand \href [0]{\begingroup \@sanitize@url \@href}%
\providecommand \@href[1]{\@@startlink{#1}\@@href}%
\providecommand \@@href[1]{\endgroup#1\@@endlink}%
\providecommand \@sanitize@url [0]{\catcode `\\12\catcode `\$12\catcode
  `\&12\catcode `\#12\catcode `\^12\catcode `\_12\catcode `\%12\relax}%
\providecommand \@@startlink[1]{}%
\providecommand \@@endlink[0]{}%
\providecommand \url  [0]{\begingroup\@sanitize@url \@url }%
\providecommand \@url [1]{\endgroup\@href {#1}{\urlprefix }}%
\providecommand \urlprefix  [0]{URL }%
\providecommand \Eprint [0]{\href }%
\providecommand \doibase [0]{https://doi.org/}%
\providecommand \selectlanguage [0]{\@gobble}%
\providecommand \bibinfo  [0]{\@secondoftwo}%
\providecommand \bibfield  [0]{\@secondoftwo}%
\providecommand \translation [1]{[#1]}%
\providecommand \BibitemOpen [0]{}%
\providecommand \bibitemStop [0]{}%
\providecommand \bibitemNoStop [0]{.\EOS\space}%
\providecommand \EOS [0]{\spacefactor3000\relax}%
\providecommand \BibitemShut  [1]{\csname bibitem#1\endcsname}%
\let\auto@bib@innerbib\@empty
\bibitem [{\citenamefont {Burr}\ \emph {et~al.}(2017)\citenamefont {Burr},
  \citenamefont {Shelby}, \citenamefont {Sebastian}, \citenamefont {Kim},
  \citenamefont {Kim}, \citenamefont {Sidler}, \citenamefont {Virwani},
  \citenamefont {Ishii}, \citenamefont {Narayanan}, \citenamefont {Fumarola},
  \citenamefont {Sanches}, \citenamefont {Boybat}, \citenamefont {Le~Gallo},
  \citenamefont {Moon}, \citenamefont {Woo}, \citenamefont {Hwang},\ and\
  \citenamefont {Leblebici}}]{burr_neuromorphic_2017}%
  \BibitemOpen
  \bibfield  {author} {\bibinfo {author} {\bibfnamefont {G.~W.}\ \bibnamefont
  {Burr}}, \bibinfo {author} {\bibfnamefont {R.~M.}\ \bibnamefont {Shelby}},
  \bibinfo {author} {\bibfnamefont {A.}~\bibnamefont {Sebastian}}, \bibinfo
  {author} {\bibfnamefont {S.}~\bibnamefont {Kim}}, \bibinfo {author}
  {\bibfnamefont {S.}~\bibnamefont {Kim}}, \bibinfo {author} {\bibfnamefont
  {S.}~\bibnamefont {Sidler}}, \bibinfo {author} {\bibfnamefont
  {K.}~\bibnamefont {Virwani}}, \bibinfo {author} {\bibfnamefont
  {M.}~\bibnamefont {Ishii}}, \bibinfo {author} {\bibfnamefont
  {P.}~\bibnamefont {Narayanan}}, \bibinfo {author} {\bibfnamefont
  {A.}~\bibnamefont {Fumarola}}, \bibinfo {author} {\bibfnamefont {L.~L.}\
  \bibnamefont {Sanches}}, \bibinfo {author} {\bibfnamefont {I.}~\bibnamefont
  {Boybat}}, \bibinfo {author} {\bibfnamefont {M.}~\bibnamefont {Le~Gallo}},
  \bibinfo {author} {\bibfnamefont {K.}~\bibnamefont {Moon}}, \bibinfo {author}
  {\bibfnamefont {J.}~\bibnamefont {Woo}}, \bibinfo {author} {\bibfnamefont
  {H.}~\bibnamefont {Hwang}},\ and\ \bibinfo {author} {\bibfnamefont
  {Y.}~\bibnamefont {Leblebici}},\ }\bibfield  {title} {\enquote {\bibinfo
  {title} {Neuromorphic computing using non-volatile memory},}\ }\href
  {https://doi.org/10.1080/23746149.2016.1259585} {\bibfield  {journal}
  {\bibinfo  {journal} {Adv. Phys. X}\ }\textbf {\bibinfo {volume} {2}},\
  \bibinfo {pages} {89--124} (\bibinfo {year} {2017})}\BibitemShut {NoStop}%
\bibitem [{\citenamefont {Li}\ \emph {et~al.}(2017)\citenamefont {Li},
  \citenamefont {Wu}, \citenamefont {Mitra},\ and\ \citenamefont
  {Wong}}]{li_resistive_2017}%
  \BibitemOpen
  \bibfield  {author} {\bibinfo {author} {\bibfnamefont {H.}~\bibnamefont
  {Li}}, \bibinfo {author} {\bibfnamefont {T.~F.}\ \bibnamefont {Wu}}, \bibinfo
  {author} {\bibfnamefont {S.}~\bibnamefont {Mitra}},\ and\ \bibinfo {author}
  {\bibfnamefont {H.-S.~P.}\ \bibnamefont {Wong}},\ }\bibfield  {title}
  {\enquote {\bibinfo {title} {Resistive {{RAM}}-{{Centric Computing}}: Design
  and {{Modeling Methodology}}},}\ }\href
  {https://doi.org/10.1109/TCSI.2017.2709812} {\bibfield  {journal} {\bibinfo
  {journal} {IEEE Trans. Circuits Syst. I}\ }\textbf {\bibinfo {volume} {64}},\
  \bibinfo {pages} {2263--2273} (\bibinfo {year} {2017})}\BibitemShut {NoStop}%
\bibitem [{\citenamefont {Sangwan}\ and\ \citenamefont
  {Hersam}(2020)}]{sangwan_neuromorphic_2020}%
  \BibitemOpen
  \bibfield  {author} {\bibinfo {author} {\bibfnamefont {V.~K.}\ \bibnamefont
  {Sangwan}}\ and\ \bibinfo {author} {\bibfnamefont {M.~C.}\ \bibnamefont
  {Hersam}},\ }\bibfield  {title} {\enquote {\bibinfo {title} {Neuromorphic
  nanoelectronic materials},}\ }\href
  {https://doi.org/10.1038/s41565-020-0647-z} {\bibfield  {journal} {\bibinfo
  {journal} {Nat. Nanotechnol.}\ }\textbf {\bibinfo {volume} {15}},\ \bibinfo
  {pages} {517--528} (\bibinfo {year} {2020})}\BibitemShut {NoStop}%
\bibitem [{\citenamefont {Waser}\ \emph {et~al.}(2009)\citenamefont {Waser},
  \citenamefont {Dittmann}, \citenamefont {Staikov},\ and\ \citenamefont
  {Szot}}]{waser_redox-based_2009}%
  \BibitemOpen
  \bibfield  {author} {\bibinfo {author} {\bibfnamefont {R.}~\bibnamefont
  {Waser}}, \bibinfo {author} {\bibfnamefont {R.}~\bibnamefont {Dittmann}},
  \bibinfo {author} {\bibfnamefont {G.}~\bibnamefont {Staikov}},\ and\ \bibinfo
  {author} {\bibfnamefont {K.}~\bibnamefont {Szot}},\ }\bibfield  {title}
  {\enquote {\bibinfo {title} {Redox-{{Based Resistive Switching Memories}} -
  {{Nanoionic Mechanisms}}, {{Prospects}}, and {{Challenges}}},}\ }\href
  {https://doi.org/10.1002/adma.200900375} {\bibfield  {journal} {\bibinfo
  {journal} {Adv. Mater.}\ }\textbf {\bibinfo {volume} {21}},\ \bibinfo {pages}
  {2632--2663} (\bibinfo {year} {2009})}\BibitemShut {NoStop}%
\bibitem [{\citenamefont {Hardtdegen}\ \emph {et~al.}(2016)\citenamefont
  {Hardtdegen}, \citenamefont {La~Torre}, \citenamefont {Zhang}, \citenamefont
  {Funck}, \citenamefont {Menzel}, \citenamefont {Waser},\ and\ \citenamefont
  {{Hoffmann-Eifert}}}]{hardtdegen_internal_2016}%
  \BibitemOpen
  \bibfield  {author} {\bibinfo {author} {\bibfnamefont {A.}~\bibnamefont
  {Hardtdegen}}, \bibinfo {author} {\bibfnamefont {C.}~\bibnamefont
  {La~Torre}}, \bibinfo {author} {\bibfnamefont {H.}~\bibnamefont {Zhang}},
  \bibinfo {author} {\bibfnamefont {C.}~\bibnamefont {Funck}}, \bibinfo
  {author} {\bibfnamefont {S.}~\bibnamefont {Menzel}}, \bibinfo {author}
  {\bibfnamefont {R.}~\bibnamefont {Waser}},\ and\ \bibinfo {author}
  {\bibfnamefont {S.}~\bibnamefont {{Hoffmann-Eifert}}},\ }\bibfield  {title}
  {\enquote {\bibinfo {title} {Internal {{Cell Resistance}} as the {{Origin}}
  of {{Abrupt Reset Behavior}} in {{HfO\textsubscript{2}}}-{{Based Devices
  Determined}} from {{Current Compliance Series}}},}\ }in\ \href
  {https://doi.org/10.1109/IMW.2016.7495280} {\emph {\bibinfo {booktitle} {2016
  {{IEEE}} 8th {{International Memory Workshop}} ({{IMW}})}}}\ (\bibinfo
  {publisher} {{IEEE}},\ \bibinfo {address} {{Paris, France}},\ \bibinfo {year}
  {2016})\ pp.\ \bibinfo {pages} {1--4}\BibitemShut {NoStop}%
\bibitem [{\citenamefont {Ib{\'a}{\~n}ez}\ \emph {et~al.}(2021)\citenamefont
  {Ib{\'a}{\~n}ez}, \citenamefont {Barrera}, \citenamefont {Maldonado},
  \citenamefont {Y{\'a}{\~n}ez},\ and\ \citenamefont
  {Rold{\'a}n}}]{ibanez_non-uniform_2021}%
  \BibitemOpen
  \bibfield  {author} {\bibinfo {author} {\bibfnamefont {M.~J.}\ \bibnamefont
  {Ib{\'a}{\~n}ez}}, \bibinfo {author} {\bibfnamefont {D.}~\bibnamefont
  {Barrera}}, \bibinfo {author} {\bibfnamefont {D.}~\bibnamefont {Maldonado}},
  \bibinfo {author} {\bibfnamefont {R.}~\bibnamefont {Y{\'a}{\~n}ez}},\ and\
  \bibinfo {author} {\bibfnamefont {J.~B.}\ \bibnamefont {Rold{\'a}n}},\
  }\bibfield  {title} {\enquote {\bibinfo {title} {Non-{{Uniform Spline
  Quasi}}-{{Interpolation}} to {{Extract}} the {{Series Resistance}} in
  {{Resistive Switching Memristors}} for {{Compact Modeling Purposes}}},}\
  }\href {https://doi.org/10.3390/math9172159} {\bibfield  {journal} {\bibinfo
  {journal} {Mathematics}\ }\textbf {\bibinfo {volume} {9}},\ \bibinfo {pages}
  {2159} (\bibinfo {year} {2021})}\BibitemShut {NoStop}%
\bibitem [{\citenamefont {Fantini}\ \emph {et~al.}(2012)\citenamefont
  {Fantini}, \citenamefont {Wouters}, \citenamefont {Degraeve}, \citenamefont
  {Goux}, \citenamefont {Pantisano}, \citenamefont {Kar}, \citenamefont {Chen},
  \citenamefont {Govoreanu}, \citenamefont {Kittl},\ and\ \citenamefont
  {Altimime}}]{fantini_intrinsic_2012}%
  \BibitemOpen
  \bibfield  {author} {\bibinfo {author} {\bibfnamefont {A.}~\bibnamefont
  {Fantini}}, \bibinfo {author} {\bibfnamefont {D.~J.}\ \bibnamefont
  {Wouters}}, \bibinfo {author} {\bibfnamefont {R.}~\bibnamefont {Degraeve}},
  \bibinfo {author} {\bibfnamefont {L.}~\bibnamefont {Goux}}, \bibinfo {author}
  {\bibfnamefont {L.}~\bibnamefont {Pantisano}}, \bibinfo {author}
  {\bibfnamefont {G.}~\bibnamefont {Kar}}, \bibinfo {author} {\bibfnamefont
  {Y.-Y.}\ \bibnamefont {Chen}}, \bibinfo {author} {\bibfnamefont
  {B.}~\bibnamefont {Govoreanu}}, \bibinfo {author} {\bibfnamefont {J.~A.}\
  \bibnamefont {Kittl}},\ and\ \bibinfo {author} {\bibfnamefont
  {L.}~\bibnamefont {Altimime}},\ }\bibfield  {title} {\enquote {\bibinfo
  {title} {Intrinsic switching behavior in {{HfO\textsubscript{2} RRAM}} by
  fast electrical measurements on novel {{2R}} test structures},}\ }in\
  \href@noop {} {\emph {\bibinfo {booktitle} {Memory {{Workshop}} ({{IMW}}),
  2012 4th {{IEEE International}}}}}\ (\bibinfo  {publisher} {{IEEE}},\
  \bibinfo {year} {2012})\ pp.\ \bibinfo {pages} {1--4}\BibitemShut {NoStop}%
\bibitem [{\citenamefont {Hennen}\ \emph {et~al.}(2019)\citenamefont {Hennen},
  \citenamefont {Bedau}, \citenamefont {Rupp}, \citenamefont {Funck},
  \citenamefont {Menzel}, \citenamefont {Grobis}, \citenamefont {Waser},\ and\
  \citenamefont {Wouters}}]{hennen_switching_2019}%
  \BibitemOpen
  \bibfield  {author} {\bibinfo {author} {\bibfnamefont {T.}~\bibnamefont
  {Hennen}}, \bibinfo {author} {\bibfnamefont {D.}~\bibnamefont {Bedau}},
  \bibinfo {author} {\bibfnamefont {J.~A.~J.}\ \bibnamefont {Rupp}}, \bibinfo
  {author} {\bibfnamefont {C.}~\bibnamefont {Funck}}, \bibinfo {author}
  {\bibfnamefont {S.}~\bibnamefont {Menzel}}, \bibinfo {author} {\bibfnamefont
  {M.}~\bibnamefont {Grobis}}, \bibinfo {author} {\bibfnamefont
  {R.}~\bibnamefont {Waser}},\ and\ \bibinfo {author} {\bibfnamefont {D.~J.}\
  \bibnamefont {Wouters}},\ }\bibfield  {title} {\enquote {\bibinfo {title}
  {Switching {{Speed Analysis}} and {{Controlled Oscillatory Behavior}} of a
  {{Cr}}-{{Doped V}}{\textsubscript{2}}{{O}}{\textsubscript{3}} {{Threshold
  Switching Device}} for {{Memory Selector}} and {{Neuromorphic Computing
  Application}}},}\ }in\ \href {https://doi.org/10.1109/IMW.2019.8739556}
  {\emph {\bibinfo {booktitle} {2019 {{IEEE}} 11th {{International Memory
  Workshop}} ({{IMW}})}}}\ (\bibinfo  {publisher} {{IEEE}},\ \bibinfo {address}
  {{Monterey, CA, USA}},\ \bibinfo {year} {2019})\ pp.\ \bibinfo {pages}
  {1--4}\BibitemShut {NoStop}%
\bibitem [{\citenamefont {Gonzalez}\ \emph {et~al.}(2020)\citenamefont
  {Gonzalez}, \citenamefont {{Maestro-Izquierdo}}, \citenamefont
  {{Jim{\'e}nez-Molinos}}, \citenamefont {Rold{\'a}n},\ and\ \citenamefont
  {Campabadal}}]{gonzalez_current_2020}%
  \BibitemOpen
  \bibfield  {author} {\bibinfo {author} {\bibfnamefont {M.~B.}\ \bibnamefont
  {Gonzalez}}, \bibinfo {author} {\bibfnamefont {M.}~\bibnamefont
  {{Maestro-Izquierdo}}}, \bibinfo {author} {\bibfnamefont {F.}~\bibnamefont
  {{Jim{\'e}nez-Molinos}}}, \bibinfo {author} {\bibfnamefont {J.~B.}\
  \bibnamefont {Rold{\'a}n}},\ and\ \bibinfo {author} {\bibfnamefont
  {F.}~\bibnamefont {Campabadal}},\ }\bibfield  {title} {\enquote {\bibinfo
  {title} {Current transient response and role of the internal resistance in
  {{HfOx}}-based memristors},}\ }\href@noop {} {\bibfield  {journal} {\bibinfo
  {journal} {Appl. Phys. Lett.}\ ,\ \bibinfo {pages} {6}} (\bibinfo {year}
  {2020})}\BibitemShut {NoStop}%
\bibitem [{\citenamefont {Maldonado}\ \emph {et~al.}(2021)\citenamefont
  {Maldonado}, \citenamefont {Aguirre}, \citenamefont {{Gonz{\'a}lez-Cordero}},
  \citenamefont {Rold{\'a}n}, \citenamefont {Gonz{\'a}lez}, \citenamefont
  {{Jim{\'e}nez-Molinos}}, \citenamefont {Campabadal}, \citenamefont
  {Miranda},\ and\ \citenamefont {Rold{\'a}n}}]{maldonado_experimental_2021}%
  \BibitemOpen
  \bibfield  {author} {\bibinfo {author} {\bibfnamefont {D.}~\bibnamefont
  {Maldonado}}, \bibinfo {author} {\bibfnamefont {F.}~\bibnamefont {Aguirre}},
  \bibinfo {author} {\bibfnamefont {G.}~\bibnamefont {{Gonz{\'a}lez-Cordero}}},
  \bibinfo {author} {\bibfnamefont {A.~M.}\ \bibnamefont {Rold{\'a}n}},
  \bibinfo {author} {\bibfnamefont {M.~B.}\ \bibnamefont {Gonz{\'a}lez}},
  \bibinfo {author} {\bibfnamefont {F.}~\bibnamefont {{Jim{\'e}nez-Molinos}}},
  \bibinfo {author} {\bibfnamefont {F.}~\bibnamefont {Campabadal}}, \bibinfo
  {author} {\bibfnamefont {E.}~\bibnamefont {Miranda}},\ and\ \bibinfo {author}
  {\bibfnamefont {J.~B.}\ \bibnamefont {Rold{\'a}n}},\ }\bibfield  {title}
  {\enquote {\bibinfo {title} {Experimental study of the series resistance
  effect and its impact on the compact modeling of the conduction
  characteristics of {{HfO}}{\textsubscript{2}}-based resistive switching
  memories},}\ }\href {https://doi.org/10.1063/5.0055982} {\bibfield  {journal}
  {\bibinfo  {journal} {Journal of Applied Physics}\ }\textbf {\bibinfo
  {volume} {130}},\ \bibinfo {pages} {054503} (\bibinfo {year}
  {2021})}\BibitemShut {NoStop}%
\bibitem [{\citenamefont {Hardtdegen}\ \emph {et~al.}(2018)\citenamefont
  {Hardtdegen}, \citenamefont {La~Torre}, \citenamefont {Cuppers},
  \citenamefont {Menzel}, \citenamefont {Waser},\ and\ \citenamefont
  {{Hoffmann-Eifert}}}]{hardtdegen_improved_2018}%
  \BibitemOpen
  \bibfield  {author} {\bibinfo {author} {\bibfnamefont {A.}~\bibnamefont
  {Hardtdegen}}, \bibinfo {author} {\bibfnamefont {C.}~\bibnamefont
  {La~Torre}}, \bibinfo {author} {\bibfnamefont {F.}~\bibnamefont {Cuppers}},
  \bibinfo {author} {\bibfnamefont {S.}~\bibnamefont {Menzel}}, \bibinfo
  {author} {\bibfnamefont {R.}~\bibnamefont {Waser}},\ and\ \bibinfo {author}
  {\bibfnamefont {S.}~\bibnamefont {{Hoffmann-Eifert}}},\ }\bibfield  {title}
  {\enquote {\bibinfo {title} {Improved {{Switching Stability}} and the
  {{Effect}} of an {{Internal Series Resistor}} in
  {{HfO}}{\textsubscript{2}}/{{TiO}}{\textsubscript{x}} {{Bilayer ReRAM
  Cells}}},}\ }\href {https://doi.org/10.1109/TED.2018.2849872} {\bibfield
  {journal} {\bibinfo  {journal} {IEEE Trans. Electron Devices}\ }\textbf
  {\bibinfo {volume} {65}},\ \bibinfo {pages} {3229--3236} (\bibinfo {year}
  {2018})}\BibitemShut {NoStop}%
\bibitem [{\citenamefont {Strachan}\ \emph {et~al.}(2013)\citenamefont
  {Strachan}, \citenamefont {Torrezan}, \citenamefont {Miao}, \citenamefont
  {Pickett}, \citenamefont {Yang}, \citenamefont {Yi}, \citenamefont
  {{Medeiros-Ribeiro}},\ and\ \citenamefont {Williams}}]{strachan_state_2013}%
  \BibitemOpen
  \bibfield  {author} {\bibinfo {author} {\bibfnamefont {J.~P.}\ \bibnamefont
  {Strachan}}, \bibinfo {author} {\bibfnamefont {A.~C.}\ \bibnamefont
  {Torrezan}}, \bibinfo {author} {\bibfnamefont {F.}~\bibnamefont {Miao}},
  \bibinfo {author} {\bibfnamefont {M.~D.}\ \bibnamefont {Pickett}}, \bibinfo
  {author} {\bibfnamefont {J.~J.}\ \bibnamefont {Yang}}, \bibinfo {author}
  {\bibfnamefont {W.}~\bibnamefont {Yi}}, \bibinfo {author} {\bibfnamefont
  {G.}~\bibnamefont {{Medeiros-Ribeiro}}},\ and\ \bibinfo {author}
  {\bibfnamefont {R.~S.}\ \bibnamefont {Williams}},\ }\bibfield  {title}
  {\enquote {\bibinfo {title} {State {{Dynamics}} and {{Modeling}} of
  {{Tantalum Oxide Memristors}}},}\ }\href
  {https://doi.org/10.1109/TED.2013.2264476} {\bibfield  {journal} {\bibinfo
  {journal} {IEEE Trans. Electron Devices}\ }\textbf {\bibinfo {volume} {60}},\
  \bibinfo {pages} {2194--2202} (\bibinfo {year} {2013})}\BibitemShut {NoStop}%
\bibitem [{\citenamefont {Kinoshita}\ \emph {et~al.}(2008)\citenamefont
  {Kinoshita}, \citenamefont {Tsunoda}, \citenamefont {Sato}, \citenamefont
  {Noshiro}, \citenamefont {Yagaki}, \citenamefont {Aoki},\ and\ \citenamefont
  {Sugiyama}}]{kinoshita_reduction_2008}%
  \BibitemOpen
  \bibfield  {author} {\bibinfo {author} {\bibfnamefont {K.}~\bibnamefont
  {Kinoshita}}, \bibinfo {author} {\bibfnamefont {K.}~\bibnamefont {Tsunoda}},
  \bibinfo {author} {\bibfnamefont {Y.}~\bibnamefont {Sato}}, \bibinfo {author}
  {\bibfnamefont {H.}~\bibnamefont {Noshiro}}, \bibinfo {author} {\bibfnamefont
  {S.}~\bibnamefont {Yagaki}}, \bibinfo {author} {\bibfnamefont
  {M.}~\bibnamefont {Aoki}},\ and\ \bibinfo {author} {\bibfnamefont
  {Y.}~\bibnamefont {Sugiyama}},\ }\bibfield  {title} {\enquote {\bibinfo
  {title} {Reduction in the reset current in a resistive random access memory
  consisting of {{NiOx}} brought about by reducing a parasitic capacitance},}\
  }\href {https://doi.org/10.1063/1.2959065} {\bibfield  {journal} {\bibinfo
  {journal} {Appl. Phys. Lett.}\ }\textbf {\bibinfo {volume} {93}},\ \bibinfo
  {pages} {033506} (\bibinfo {year} {2008})}\BibitemShut {NoStop}%
\bibitem [{\citenamefont {Lu}\ \emph {et~al.}(2012)\citenamefont {Lu},
  \citenamefont {Noman}, \citenamefont {Chen}, \citenamefont {Salvador},
  \citenamefont {Bain},\ and\ \citenamefont
  {Skowronski}}]{lu_elimination_2012}%
  \BibitemOpen
  \bibfield  {author} {\bibinfo {author} {\bibfnamefont {Y.~M.}\ \bibnamefont
  {Lu}}, \bibinfo {author} {\bibfnamefont {M.}~\bibnamefont {Noman}}, \bibinfo
  {author} {\bibfnamefont {W.}~\bibnamefont {Chen}}, \bibinfo {author}
  {\bibfnamefont {P.~A.}\ \bibnamefont {Salvador}}, \bibinfo {author}
  {\bibfnamefont {J.~A.}\ \bibnamefont {Bain}},\ and\ \bibinfo {author}
  {\bibfnamefont {M.}~\bibnamefont {Skowronski}},\ }\bibfield  {title}
  {\enquote {\bibinfo {title} {Elimination of high transient currents and
  electrode damage during electroformation of {{TiO}} {\textsubscript{2}}-based
  resistive switching devices},}\ }\href
  {https://doi.org/10.1088/0022-3727/45/39/395101} {\bibfield  {journal}
  {\bibinfo  {journal} {J. Phys. Appl. Phys.}\ }\textbf {\bibinfo {volume}
  {45}},\ \bibinfo {pages} {395101} (\bibinfo {year} {2012})}\BibitemShut
  {NoStop}%
\bibitem [{\citenamefont {Sharma}\ \emph {et~al.}(2014)\citenamefont {Sharma},
  \citenamefont {Noman}, \citenamefont {Abdelmoula}, \citenamefont
  {Skowronski},\ and\ \citenamefont {Bain}}]{sharma_electronic_2014}%
  \BibitemOpen
  \bibfield  {author} {\bibinfo {author} {\bibfnamefont {A.~A.}\ \bibnamefont
  {Sharma}}, \bibinfo {author} {\bibfnamefont {M.}~\bibnamefont {Noman}},
  \bibinfo {author} {\bibfnamefont {M.}~\bibnamefont {Abdelmoula}}, \bibinfo
  {author} {\bibfnamefont {M.}~\bibnamefont {Skowronski}},\ and\ \bibinfo
  {author} {\bibfnamefont {J.~A.}\ \bibnamefont {Bain}},\ }\bibfield  {title}
  {\enquote {\bibinfo {title} {Electronic {{Instabilities Leading}} to
  {{Electroformation}} of {{Binary Metal Oxide}}-based {{Resistive
  Switches}}},}\ }\href {https://doi.org/10.1002/adfm.201400461} {\bibfield
  {journal} {\bibinfo  {journal} {Adv. Funct. Mater.}\ }\textbf {\bibinfo
  {volume} {24}},\ \bibinfo {pages} {5522--5529} (\bibinfo {year}
  {2014})}\BibitemShut {NoStop}%
\bibitem [{\citenamefont {Meng}\ \emph {et~al.}(2020)\citenamefont {Meng},
  \citenamefont {Zhao}, \citenamefont {Xu}, \citenamefont {Goodwill},
  \citenamefont {Bain},\ and\ \citenamefont
  {Skowronski}}]{meng_temperature_2020}%
  \BibitemOpen
  \bibfield  {author} {\bibinfo {author} {\bibfnamefont {J.}~\bibnamefont
  {Meng}}, \bibinfo {author} {\bibfnamefont {B.}~\bibnamefont {Zhao}}, \bibinfo
  {author} {\bibfnamefont {Q.}~\bibnamefont {Xu}}, \bibinfo {author}
  {\bibfnamefont {J.~M.}\ \bibnamefont {Goodwill}}, \bibinfo {author}
  {\bibfnamefont {J.~A.}\ \bibnamefont {Bain}},\ and\ \bibinfo {author}
  {\bibfnamefont {M.}~\bibnamefont {Skowronski}},\ }\bibfield  {title}
  {\enquote {\bibinfo {title} {Temperature overshoot as the cause of physical
  changes in resistive switching devices during electro-formation},}\ }\href
  {https://doi.org/10.1063/5.0010882} {\bibfield  {journal} {\bibinfo
  {journal} {J. Appl. Phys.}\ }\textbf {\bibinfo {volume} {127}},\ \bibinfo
  {pages} {235107} (\bibinfo {year} {2020})}\BibitemShut {NoStop}%
\bibitem [{DS1(2001)}]{DS1808}%
  \BibitemOpen
  \href {https://datasheets.maximintegrated.com/en/ds/DS1808.pdf} {\emph
  {\bibinfo {title} {DS1808 Dual Log Digital Potentiometer}}},\ \bibinfo
  {organization} {Maxim Integrated} (\bibinfo {year} {2001}),\ \bibinfo {note}
  {archived at https://perma.cc/VK7V-8MSG}\BibitemShut {NoStop}%
\bibitem [{THS(2015)}]{THS3091}%
  \BibitemOpen
  \href {https://www.ti.com/lit/ds/symlink/ths3091.pdf} {\emph {\bibinfo
  {title} {THS309x High-voltage, Low-distortion, Current-feedback Operational
  Amplifiers}}},\ \bibinfo {organization} {Texas Instruments} (\bibinfo {year}
  {2015}),\ \bibinfo {note} {archived at
  https://perma.cc/PQX3-RUZD}\BibitemShut {NoStop}%
\bibitem [{\citenamefont {Hennen}\ \emph {et~al.}(2018)\citenamefont {Hennen},
  \citenamefont {Bedau}, \citenamefont {Rupp}, \citenamefont {Funck},
  \citenamefont {Menzel}, \citenamefont {Grobis}, \citenamefont {Waser},\ and\
  \citenamefont {Wouters}}]{hennen_forming-free_2018}%
  \BibitemOpen
  \bibfield  {author} {\bibinfo {author} {\bibfnamefont {T.}~\bibnamefont
  {Hennen}}, \bibinfo {author} {\bibfnamefont {D.}~\bibnamefont {Bedau}},
  \bibinfo {author} {\bibfnamefont {J.~A.~J.}\ \bibnamefont {Rupp}}, \bibinfo
  {author} {\bibfnamefont {C.}~\bibnamefont {Funck}}, \bibinfo {author}
  {\bibfnamefont {S.}~\bibnamefont {Menzel}}, \bibinfo {author} {\bibfnamefont
  {M.}~\bibnamefont {Grobis}}, \bibinfo {author} {\bibfnamefont
  {R.}~\bibnamefont {Waser}},\ and\ \bibinfo {author} {\bibfnamefont {D.~J.}\
  \bibnamefont {Wouters}},\ }\bibfield  {title} {\enquote {\bibinfo {title}
  {Forming-{{Free Mott}}-{{Oxide Threshold Selector Nanodevice Showing}}
  s-{{Type NDR}} with {{High Endurance}} ({$>$} 10{\textsuperscript{12}}
  {{Cycles}}), {{Excellent V}}{\textsubscript{th}} {{Stability}} ({$<$} 5\%),
  {{Fast}} ({$<$} 10 ns) {{Switching}}, and {{Promising Scaling
  Properties}}},}\ }in\ \href@noop {} {\emph {\bibinfo {booktitle}
  {{{IEDM}}}}}\ (\bibinfo {address} {{San Francisco, CA, USA}},\ \bibinfo
  {year} {2018})\ pp.\ \bibinfo {pages} {37.5.1--37.5.4}\BibitemShut {NoStop}%
\bibitem [{\citenamefont {Hennen}\ \emph {et~al.}(2021)\citenamefont {Hennen},
  \citenamefont {Wichmann}, \citenamefont {Elias}, \citenamefont {Lille},
  \citenamefont {Mosendz}, \citenamefont {Waser}, \citenamefont {Wouters},\
  and\ \citenamefont {Bedau}}]{hennen_current-limiting_2021}%
  \BibitemOpen
  \bibfield  {author} {\bibinfo {author} {\bibfnamefont {T.}~\bibnamefont
  {Hennen}}, \bibinfo {author} {\bibfnamefont {E.}~\bibnamefont {Wichmann}},
  \bibinfo {author} {\bibfnamefont {A.}~\bibnamefont {Elias}}, \bibinfo
  {author} {\bibfnamefont {J.}~\bibnamefont {Lille}}, \bibinfo {author}
  {\bibfnamefont {O.}~\bibnamefont {Mosendz}}, \bibinfo {author} {\bibfnamefont
  {R.}~\bibnamefont {Waser}}, \bibinfo {author} {\bibfnamefont {D.~J.}\
  \bibnamefont {Wouters}},\ and\ \bibinfo {author} {\bibfnamefont
  {D.}~\bibnamefont {Bedau}},\ }\bibfield  {title} {\enquote {\bibinfo {title}
  {Current-limiting amplifier for high speed measurement of resistive switching
  data},}\ }\href@noop {} {\bibfield  {journal} {\bibinfo  {journal} {Rev Sci
  Instrum}\ ,\ \bibinfo {pages} {7}} (\bibinfo {year} {2021})}\BibitemShut
  {NoStop}%
\end{thebibliography}%

\end{document}